[Public]

# Zen-Attention: A Compiler Framework for Dynamic Attention Folding on AMD NPUs

Aadesh Deshmukh
*AMD*
*SanJose, CA, USA*
*Aadesh.Deshmukh@amd.com*

Venkata Yaswanth Raparti
*AMD*
*SanJose, CA, USA*
*Yaswanth.Raparti@amd.com*

Samuel Hsu
*AMD*
*SanJose, CA, USA*
*Samuel.Hsu@amd.com*

*Abstract*—**Transformer-based deep learning models are increasingly deployed on energy, and DRAM bandwidth constrained devices such as laptops and gaming consoles, which presents significant challenges in meeting the latency requirements of the models. The industry is turning to neural processing units (NPUs) for superior performance-per-watt (perf/watt); however, efficiently mapping dynamic attention layers to the NPUs remains a challenging task. For optimizing perf/watt, AMD XDNA NPUs employ software managed caches and share system memory with host. This requires substantial engineering effort to unlock efficient tiling, buffer allocation, and data movement to extract the maximum efficiency from the device. This paper introduces Zen-Attention, a framework that optimizes DRAM bandwidth utilization in the attention layer of models by systematically exploring the complex design space of layer folding, tiling, and data-movement on the interconnect, and the tensor layouts to come up with an optimal solution. Our evaluation includes comparative analysis of end-to-end model latency and specific attention latency in each model. We demonstrate how the framework enhances mapping capabilities by varying input dimensions, which require padding and masking in the attention block. For representative transformer models, the Zen-Attention Framework achieves up to 4x improvement in the latency of the attention block and up to 32% improvement in end-to-end network latency compared to the baseline Unfolded- approaches.**

## I. INTRODUCTION

The attention mechanism, fundamental to modern transformer architectures, is inherently characterized by its memory-bound computational profile. This phenomenon has been extensively documented with roofline plots that categorize deep neural network operations into two primary computational regimes: compute-bound and memory-bound operations. Attention mechanisms consistently fall into the memory-bound category across diverse hardware architectures due to low operational intensity [6]. This classification stems from several key factors: large activation tensors, reduction operations, and the prevalence of element-wise operations. The memory-bound nature of attention block presents significant performance challenges, as these operations are limited by data movement rather than computational throughput. To improve the latency, common optimizations focus on two primary techniques: maximizing data reuse [5, 6] and restructuring computations to leverage throughput-intensive kernel instructions.

AMD XDNA™ NPU [7] is a reconfigurable hardware architecture that provides an alternative way to accelerate ML workload instead of GPUs to minimize energy consumption. They provide explicit control over memory hierarchies to programmers by replacing traditional cache mechanisms with scratch buffers, resulting in significantly lower energy consumption compared to conventional processors that continuously execute energy-intensive cache line fills, associative searches, replacement policies and speculative fetching. The trade-off is that memory optimization becomes the programmer's responsibility, as developers must explicitly manage data movement patterns between on-chip and off-chip memory to achieve optimal performance. While this approach requires more sophisticated programming, it enables precise control over both data movement and energy usage, making NPUs attractive for applications where latency and power efficiency are paramount. The memory hierarchy contains three levels: L1 (core-tile memory), L2 (shared memory, or memory-tile) and L3 is the host memory subsystem. L1 and L2 serve as scratchpad memory, while L3 serves as host memory on DRAM. NPU shares the L3 memory with the rest of the system, including the CPU and GPU. NPU cores are organized spatially in a 2D-array layout, with each column sharing L2 memory. Core tiles communicate with the other tiles through an interconnection network. Each tile has a DMA engine to facilitate data movement from stream-to-memory and memory-to-stream. *Dedicated cascade streams connect cores within the array, to support spatial reduction across the cores.* Shim-tiles have ports to access DRAM or the last-level cache of the host memory sub-system.

The most common approach for mapping attention layers onto the NPU architecture follows a layer-by-layer strategy. Optimized kernels are developed for each layer (GeMM, SoftMax, Add) to execute them independently by reading the inputs from host memory and writes the output back to the host memory. A layer (machine learning operator) is spatially *tiled* into sub-volumes and mapped onto specific compute cores. Temporal iterations handle cases where the size of the output exceed the compute and memory capacity of the NPU array. The design operates under constraints of L1 and L2 memory capacities. The layer output is sent in subvolumes to L3 (part of host CPU memory hierarchy). More optimized mapping schemes save the layer output in L1/L2 memory and optimize inter-layer roundtrip latency. This technique is called *layer folding.* Layer folding is constrained by several factors such as L1 and L2 sizes, and the locality of intermediate output on the array. Hence a comprehensive tiling and buffer allocation framework is necessary for optimized layer folding on the NPUs.

The NPUs are highly sensitive to the data movement due to their lower memory bandwidth-to-compute ratio compared to GPUs. For example, AMD Ryzen™ AI 9 HX platforms allocate only 60 GB/s for the NPU which is less than half of the bandwidth allocated for the GPU (130 GB/s). In contrast, discrete GPUs achieve their throughput via dedicated high bandwidth memory and low context switch overhead between wavefronts. NPU compute cores have VLIW pipeline with special vector and matrix multiplication (MatMul) compute units. Each tile has an explicitly programmable DMA engine with multiple channels, to achieve parallelism.

## II. PROBLEMS WITH MULTI-HEAD-ATTENTION

Current NPU research predominantly focuses on offloading high-throughput intensive GEMMs on accelerators while executing remaining operations on the CPUs, introducing significant NPU-to-CPU I/O and context-switch overhead [1-4]. Existing mechanisms that attempt layer-by-layer execution result in excessive DRAM roundtrips for activation data, which constitutes the primary source of memory bottleneck in the attention layers on the NPUs. Additionally, attention layers require data transformations such as Transpose before or after MatMul which requires special handling in the cores, or on CPU. The inefficiencies grow exponentially if the inputs and the outputs are not divisible by the compute granularities of the cores which require extensive padding. Most activation data transformations require DRAM round trips, which is inefficient and further exacerbates the memory-bottleneck problem. Kao et al. has proposed to improve the performance and efficiency of different individual operators such as Query*Key ($Q*K^T$), SoftMax (*SM*), and SoftMax*V (*SM*V*) through kernel folding [6]. This minimizes the roundtrip latency to DRAM for each operator as shown in figure 1 by reusing the data in L1. However, this is limited by the L1 size of each core tile that executes the kernel. On the NPU, we can leverage the cascade streams across the cores as shown in figure 2 to extend the scope for folding.

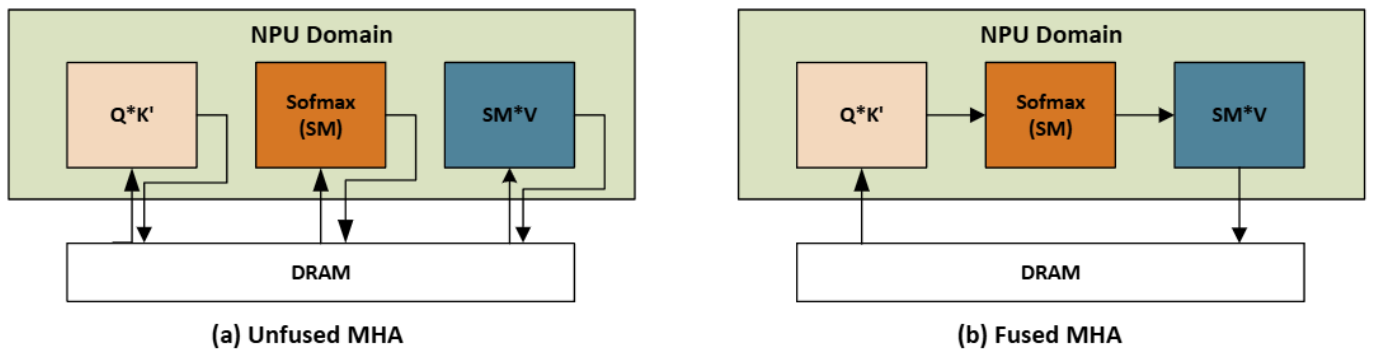

**Figure 1. Implementation of (a) Unfolded MHA (b) Folded MHA.**

Further, modern language models leverage three distinct attention architectures that have emerged to solve the memory bottleneck crisis. In multi-head attention (MHA) each head maintains separate query, key, and value matrices. Memory requirements scale quadratically with sequence length and head count. Multi-query attention (MQA) shares single key-value heads across query heads, reducing the memory requirement. Group query attention (GQA) is a middle ground where each group shares key-value representations internally. Memory efficiency improves while performance remains competitive. Different variants of attention blocks require different kernel implementations, tiling, memory allocation, and data movement orchestration. Additionally, in some models, *Masks* and *Biases* are injected into the attention block which further increases the memory requirement for L1 folding. To address these problems, we introduce a novel compiler framework that maps the attention block as a single folded layer on the NPU array minimizing the reads and writes to the DRAM. The following are our main contributions.

1. A novel hardware-aware graph optimization heuristic to fuse different layers within the attention block.
2. An efficient *folded attention* based tiler to derive the maximum utilization on the NPU by leveraging L1 folding and spatial reduction.
3. An intelligent way to handle data transformations such as input/output transpose, padding, and de-padding as a part of folding and reduce additional DRAM traffic.

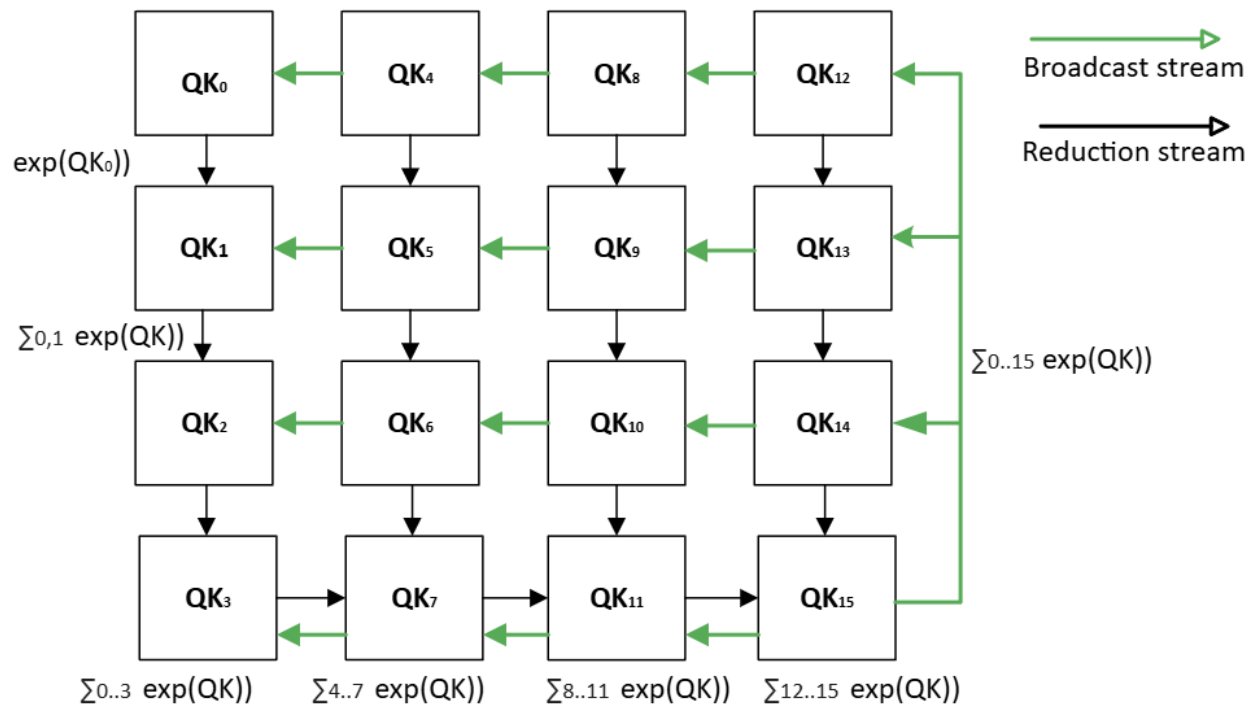

**Figure 2. Demonstration of spatial reduction in SoftMax layer.**

## III. ZEN-ATTENTION FRAMWORK

### A. *Hardware Aware Graph Optimization*

In this stage we attempt to optimize the graph. At first, we analyze the ONNX graph and perform a two-step attention folding and graph transformation processes. In the first step the folding of the following chain of operations is attempted.

$$A = Q * K^T \quad (1)$$

$$A = A + B + M \ (optional) \quad (2)$$

$$SM_{out} = SoftMax(A) \quad (3)$$

$$Z = SM_{out} * V \quad (4)$$

Specific patterns indicating distinct operators in the attention blocks are used to match and verify if the input shapes can fit into L1 memory for layer folding. Graph optimizer invokes the *Tiler* to check whether a chain of operators can be folded into a single node representing the *folded attention* operation. The Tiler returns a *folding_level*, to indicate how many layers in the attention block can be folded. The second step is batching the attention nodes across different heads to create Batched attention nodes and classify them into MHA, GQA or MQA operations. Then, we check for the optional presence of the *mask* and the *bias* addition and try to optimize them. While batching, we detect the nodes that can be executed together and generate a sequence to facilitate removing nodes of individual *heads* in the graph.

### B. *Tiling*

The tiler takes Tensor dimensions of *Q*, *K*, *V*, *B*, *M* tensors and L1 size as inputs and finds the subvolumes $\{S_q\, S_k\, S_v\, S_b\, S_m\}$ that can fit into L1 without violating buffer allocation constraints (prescribed by the kernels). The tiling logic first creates a list of *tilings* that are valid. The list is sorted in decreasing order to maximize the size of *Sq*. The tiling logic then iterates over the tilings until it finds suitable subvolumes that fit in L1. If *K, V* are pinned to L1, *Sq* is unrolled across the cores within the same column. Similarly, batch size (*Heads*) can also be unrolled across the columns. However, when *K, V, and M* are spatially split batch size is temporally iterated as we leverage spatial reduction across the columns as shown in figure 2. We further define a metric called *folding_level* which determines how many layers are folded. *Folding_level*=3 indicates the folding of all the layers in the attention block. In *folding_level*=2 we avoid fusing the last MatMul (*SM * V*) due to the lack of L1 size, which is still better performing than *folding_level*=1 which indicates unfolded attention.

### C. *Transpose.*

Transpose is treated as a separate operator on host or on the NPU which is expensive in terms of compute and DRAM access overheads. Further, when the transpose operator is folded with the attention block it requires additional L1 buffers which limit the feasibility of fusing the attention block with long context lengths. To address the above issues, we propose a novel *Folding-Preserving Transpose* mechanism that maintains folding integrity while efficiently handling transpose operations. This approach combines DMA-based *block-transpose* operations on L2 buffer data, and *intra-block* level transpose operations on L1 buffer data using a specialized MatMul kernel. The key constraint for data transformation using DMA is the minimum granularity of traversal. However, as the minimum stride supported on DMA is 4B, we cannot use DMA to truly transpose a tensor. To overcome this problem, we transpose the data in blocks of 8×8 granularity and transpose the blocks inside the kernel using *SHUFFLE* intrinsics at register level before performing the MatMul. This optimized kernel is called *Transposed-MatMul*. This hybrid approach eliminates the need for separate transpose kernel and the L1 buffer, thereby achieving near-optimal performance.

[Public]

### D. Padding

The compute-kernels that process the tiled inputs on the cores have minimum required dimensions on each axis prescribed by the lane widths of *VMAC* and *VMUL* instructions. Further, the loops within compute intensive MatMul and SoftMax kernels are unrolled to achieve software pipelining, adding strict lower boundaries on tensor dimensions. It is not uncommon to see odd dimensions in the input and output tensors of the attention block. To meet the kernel requirements, the tensors are padded to the next multiples of kernel granularities. In AMD-XDNA2 NPU, L2 *MM2S*(read) DMA channels offer padding across {*D0, D1, D2*} dimensions. We can leverage this feature to pad the odd shapes to the multiples of kernel granularities. In rare cases where DMA padding is not feasible, we rely on the previous layer to pad the corresponding output tensor to meet the requirements, hence eliminating the need for a special Pad operator.

## IV. Experiments

We conducted experiments with a set of 5 models which represent recent language models with attention blocks. We validated these layers and the models on AMD Ryzen AI 9 HX 370 machines with a 32 core NPU arranged in a 4x8 grid delivering 50 TOPs. The NPU array has 64KB per L1, and 512KB per L2 tile. The processor subsystem has 12 Zen 5 series CPU cores. Both the CPU and the NPU are connected to a shared DRAM which delivers a combined read-write bandwidth of ~60 GB/s. Our framework uses precompiled kernels of the attention layers, unlike [6], by leveraging the inherent separation between compute and dataflow resources of the NPU architecture. Table 1 shows some important characteristics of the models. Each model has 1 or 2 unique shapes of the attention blocks. Some models require padding to their input dimensions to meet kernel granularities. All the models require Transpose for the *K* input which is folded by the Zen-attention Framework. Figure 4 shows the improvement in latency between folded and unfolded attention blocks across different models with the optimizations proposed in the Zen-Attention Framework. Intuitively, models with larger sequence *(Q)* and context lengths *(K)* achieve higher speed-up in comparison with models with smaller *Q, K*. This is correlated with the DRAM bandwidth requirements of each model. *Folded attention* minimizes the DRAM roundtrip of the bandwidth intensive layers of the attention block which results in up to ~4× lower latency. For smaller dimensions, the improvement is not that significant as the compute time overshadows the bandwidth requirements. Furthermore, smaller shapes leave some NPU cores unused which diminishes the compute efficiency making the attention block compute bound. Even in compute bound scenarios, *folded attention* shows ~8% lower latency compared to Unfolded, and lower DRAM bandwidth utilization, which benefits applications executed simultaneously on the memory constrained system.

| Model | Attn shapes | Mask | Num Heads | Padding |
|---|---|---|---|---|
| ViT-base-patch | 1 | × | 12 | ✓ |
| CLIP-patch32 | 2 | ✓ | 80, 12 | ✓ |
| CLIP-patch16 | 2 | ✓ | 80, 12 | ✓ |
| CLIP-Laion | 2 | ✓ | 80, 12 | ✓ |
| BERT | 1 | ✓ | 12 | × |

**Table 1: Models evaluated using Zen-Attention Framework.**

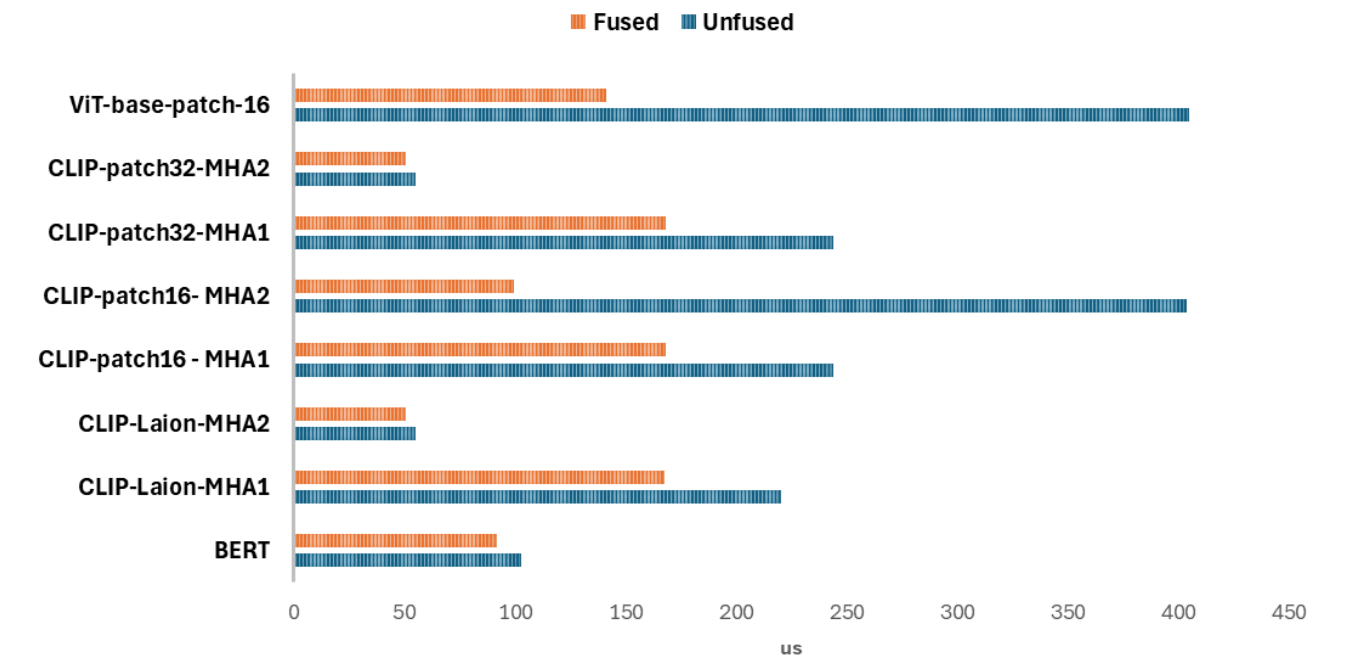

**Figure 4: Comparison of attention block latency between Folded and Unfolded approaches across different networks.**

Figure 5 shows the improvement in end-to-end latency when Zen-Attention Framework is employed to map the attention block. With *folded attention*, there is up to 32% improvement in network latency in ViT-base-patch-16 model which demonstrates the impact of bandwidth bottleneck induced by the attention layers. The minimum latency noticed is 1.4% with BERT where the attention block is not the major source of performance bottleneck. However, the applications that run concurrently are expected to benefit with reduced DRAM bandwidth utilization.

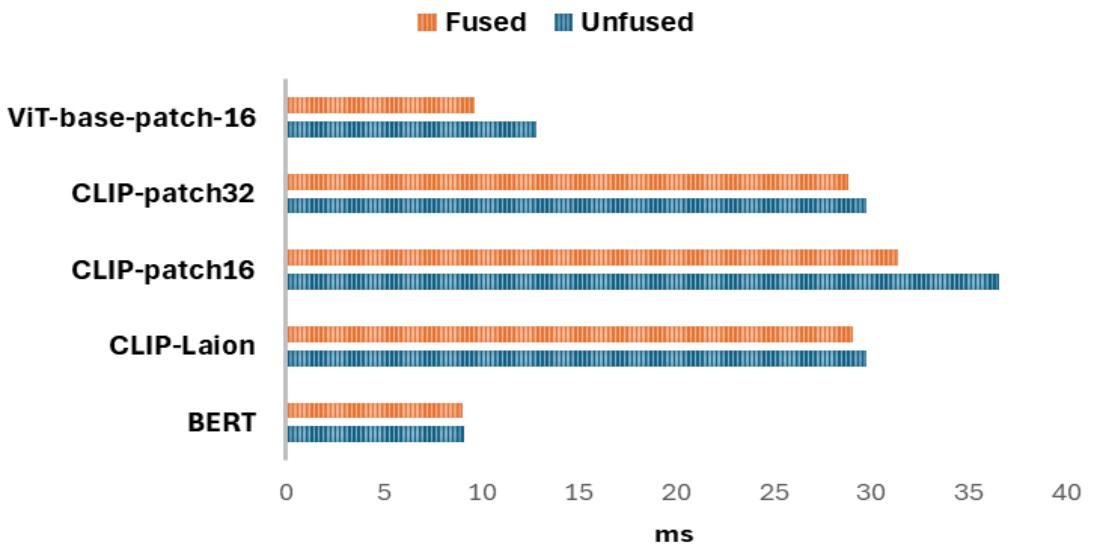

**Figure 5: Comparison of end-to-end network latency between Folded and Unfolded attention blocks across different networks.**